\documentclass[aps,prb,twocolumn,showpacs,superscriptaddress]{revtex4}
\usepackage{times,xspace}
\usepackage{amsbsy,amssymb,amsmath,bm}
\usepackage{graphicx,color,epsfig,rotate}
\usepackage{fancyheadings}

\begin{document}
\title{ELECTROMECHANICAL PROPERTIES OF MULTI-DOMAIN FERROELECTRICS}
\author{Rajeev Ahluwalia}
\author{Turab Lookman}
\author{Avadh Saxena}
\affiliation{Theoretical Division, Los
Alamos
National Laboratory,
Los Alamos, New Mexico, 87545 }
\author{Wenwu Cao}
\affiliation{Materials Research Institute and Department of Mathematics, Pennsylvania State University,
Pennsylvania 16802 }
\date{\today}
\begin{abstract}
We study theoretically the influence of the underlying domain 
microstructure on
the electromechanical properties of ferroelectrics. Our
calculations are based on a continuum approach that incorporates
the long-range elastic and electrostatic interactions. The theory
is used to simulate the piezoelectric properties of a two
dimensional model ferroelectric crystal. Simulation results
indicate that the electromechanical response of the
ferroelectric is strongly dependent on the domain
microsctructure, including domain walls. This is particularly
true for the case when an electric field is applied along a
non-polar direction. 
\end{abstract}
\maketitle
Ferroelectrics are the best piezoelectric materials that can convert electrical
energy into mechanical energy and vice versa \cite{glass}. This
electromechanical property
arises due to the coupling of the spontaneous polarization with
lattice strain. 
Many devices such as ultrasonic transducers and piezoelectric actuators make use of this
property\cite{uchino}. Recently, there has been considerable
interest in this field due to the observation of a giant
piezoelectric response when the applied field is along a non-polar
direction \cite{shrout,wada}.
 It is believed that this
``superpiezoelectric" response is due to the symmetry change caused
by a rotation of the polarization
 towards the direction of the applied field\cite{cohen}.
Domain configurations produced by the non-polar direction field are termed {\it engineered domains}.
There are also a large number of domain walls between the degenerate variants which affect
the piezoelectric property.
It is important to understand the role played by such domain microstructures on the
piezoelectric response.

Electromechanical properties of ferroelectrics have been studied
theoretically using first-principle calculations
\cite{david,cohen,nasai}. A continuum Landau theory describing a
single domain or homogeneous state has been used to study the
electromechanical properties of $BaTi{O_3}$ as function of
temperature and electric field direction \cite{bell}. Although
such calculations provide valuable insights into the physics of
the polarization-strain coupling, they do not describe
inhomogeneities due to domains and domain walls. In this Letter,
we study the behavior of domain patterns under applied electric field and
investigate how the microstructural evolution influences the
average electromechanical response of ferroelectric materials.

Our approach is to use a
time-dependent Ginzburg-Landau model \cite{lqc,rawc} with long-range elastic and electrostatic effects.
We
restrict ourselves to a $2D$ ferroelectric transition to
illustrate the basic principles  and use parameters from a model for
$BaTi{O_3}$ in our calculations \cite{bell}. We find that the electromechanical response is highly orientation
and microstructure dependent. When the field is applied along one of the polar
directions, domain switching results in higher strains compared to the single domain state.  For fields applied along a
non-polar direction, the domain walls serve as nucleation sites for a field induced structural phase transition.

The free-energy functional for a $2D$ ferroelectric system is written as
$F=F_{l} + F_{em} +F_{es}$.
Here $F_{l}$ is the local free energy \cite{bell} that describes the
ferroelectric transformation and is given as
\begin{eqnarray}
&F_{l}&=\int d\vec{r}\bigg\{{\alpha_1}({P_x}^2+{P_y}^2)
+{\alpha_{11}}({P_x}^4+{P_y}^4)\nonumber\\
&+&{\alpha_{12}}{P_x}^2{P_y}^2
+{\alpha_{111}}({P_x}^6+{P_y}^6)\nonumber\\
&+&{\alpha_{112}}({P_x}^2{P_y}^4
+{P_x}^4{P_y}^2)
-{E_x}{P_x}-{E_y}{P_y}\nonumber\\
&+&{{g_1}\over{2}}( {P_{x,x}}^2
+{P_{y,y}}^2)
+
{{g_2}\over{2}}({P_{x,y}}^2+
{P_{y,x}}^2)\nonumber\\
&+&
{g_3}{P_{x,x}}{P_{y,y}}
\bigg\} , 
\end{eqnarray}
 where $P_x$ and $P_y$ are the polarization components.
The free energy coefficients $\alpha_1, \alpha_{11},...,\alpha_{112}$ 
determine the ferroelectric phase and the gradient
coefficients $g_1$, $g_2$ and $g_3$ are a measure of domain wall energies. $E_x$ and $E_y$ are the components of an external electric field.
Elastic properties are studied by
using the strains $\eta_1=\eta_{xx}+\eta_{yy}$
, $\eta_2=\eta_{xx}-\eta_{yy}$ and
$\eta_3=\eta_{xy}$, where $\eta_{ij}$ is the
linearized strain tensor defined as
 $\eta_{ij}=(u_{i,j}+u_{j,i})/2$ $(i,j=x,y)$, $u_i$ being the components of the displacement
vector.
The electromechanical coupling is described in terms of these strain variables as
the free energy
$F_{em}=\lambda\int d\vec{r}[\{\eta_1-{Q_1}({P_x}^2+{P_y}^2)\}^2+
\{\eta_2-{Q_2}({P_x}^2-{P_y}^2)\}^2
+\{\eta_3-{Q_3}{P_x}{P_y}\}^2]$.
Here $Q_1$, $Q_2$ and $Q_3$ are obtained from the
electrostrictive constants of the material as $Q_1=Q_{11}+Q_{12}$,
$Q_2=Q_{11}-Q_{12}$ and
$Q_3=Q_{44}$ (electrostrictive constants describe coupling between strains and polarization as
$\eta_{xx}=Q_{11}{P_x}^2+Q_{12}{P_y}^2$,
$\eta_{yy}=Q_{11}{P_y}^2+Q_{12}{P_x}^2$ and
$\eta_{xy}=Q_{44}{P_x}{P_y}$). Notice that the free energy $F_{em}$ vanishes 
for
a homogeneous state as the homogeneous strains in equilibrium are given by
${\eta_1}^{e}={Q_1}({P_x}^2+{P_y}^2)$,
${\eta_2}^{e}={Q_2}({P_x}^2-{P_y}^2)$,
and ${\eta_3}^{e}={Q_3}{P_x}{P_y}$. However, this free energy does not vanish for an inhomogeneous
state. For an inhomogeneous state,
the strains $\eta_1$, $\eta_2$ and
$\eta_3$ are related to each other by the elastic compatibility constraint \cite{love}
${{\nabla}^2}{\eta_1}-({{\partial^2}\over{\partial x^2}}
-{{\partial^2}\over{\partial y^2}}){\eta_2}-
{{\partial^2}\over{ {\partial x}{\partial y} }}{\eta_3}=0.$
Using this relation, the strain $\eta_1$ can be eliminated from $F_{em}$ resulting in a nonlocal interaction between the strains involving $\eta_2$ and $\eta_3$. Using the equilibrium strains defined by
${\eta_2}^{e}$ and
${\eta_3}^{e}$, the electromechanical free energy can be written as
\begin{equation}
F_{em}=\lambda\int d \vec{k}|{C_2(\vec{k})}\Gamma_2(\vec{k})
+{C_3(\vec{k})}\Gamma_3(\vec{k})
-\Gamma_1(\vec{k})|^2,
\end{equation}
 where the $\vec{k}=0$ mode has been excluded from the above integral.
The constant $\lambda$ is the strength of this nonlocal interaction and hence it influences the underlying
microstructure.
The quantities $\Gamma_1(\vec{k})$, $\Gamma_2(\vec{k})$ and $\Gamma_3(\vec{k})$ are respectively the Fourier
transforms of $Q_1({P_x}^2+{P_y}^2)$
, $Q_2({P_x}^2-{P_y}^2)$ and
$Q_3{P_x}{P_y}$,
$C_{2}=({k_x}^2-{k_y}^2)/
({k_x}^2+{k_y}^2)$ and
 $C_{3}={k_x}{k_y}/
({k_x}^2+{k_y}^2)$ are the orientation dependent kernels. The
electrostatic contribution to the free energy is calculated by
considering the depolarization energy \cite{brat} $F_{es}=-\mu\int d\vec{r}
\{\vec{E_d}\cdot\vec{P}
+{\epsilon_0}({\vec{E_d}\cdot\vec{E_d}}/{2})\}$, where
$\vec{E_d}$ is the internal depolarization field due to the
dipoles and $\mu$ is the strength of this interaction. The field
$\vec{E_d}$ can be calculated from an underlying potential as
$\vec{E_d}=-\vec{\nabla}\phi$. If we assume that there is no free
charge in the system, then $\vec{\nabla}\cdot\vec{D}=0$, where
$\vec{D}$ is the electric displacement vector
 defined by $\vec{D}={\epsilon_0}\vec{E_d}+\vec{P}$. This equation gives rise to the
constraint
$-{\epsilon_0}{\nabla}^2{\phi}+\vec{\nabla}\cdot\vec{P}=0$. The
potential $\phi$ is eliminated from the free energy $F_{es}$ using
the above constraint to express $F_{es}$ in Fourier space as
\begin{equation}
F_{es}=(\mu/2{\epsilon_0})\int d\vec{k}|\hat{k_x}{P_x}(\vec{k})
+\hat{k_y}{P_y}(\vec{k})
|^2.
\end{equation}
The above integral excludes the homogeneous $\vec{k}=0$ mode
which means that the homogeneous depolarization field due to
surface charges has been neglected. The total energy is defined as
$F=F_{l}+F_{em}+F_{es}$ with two additional constants, i.e.
$\lambda$ and $\mu$ are essential for the description of
multi-domain states.

The dynamics of the polarization fields is given by the relaxational 
time-dependent Ginzburg-Landau equations
${{\partial P_i}\over{\partial t}}=-
{\gamma}{{\delta F}\over{\delta P_i}}$,
where $\gamma$ is a dissipation coefficient and $i=x,y$ represents the polarization components.
We first introduce rescaled variables defined as
$u=P_x/{P_0}$,
$v=P_y/{P_0}$, $\vec{\zeta}=\vec{r}/\delta$ and $t^*=\gamma|\alpha_1(T_0)|{t}$, where
$T_0$ is a fixed temperature.
In this work, we use the parameters \cite{bell} for $BaTi{O_3}$ for the local part of the free energy
$F_{l}$. The parameters which can be dependent on the temperature $T$ are:
$\alpha_1=3.34\times{10^5}(T-381)$ VmC$^{-1}$,
$\alpha_{11}=4.69\times{10^6}(T-393) -2.02\times{10^8}$ 
Vm$^{5}$C$^{-3}$,
$\alpha_{111}=-5.52\times{10^7}(T-393)
+2.76\times{10^9}$ Vm$^{9}$C$^{-5}$,
$\alpha_{12}=3.23\times{10^8}$ Vm$^{5}$C$^{-3}$ and
$\alpha_{112}=4.47\times{10^9}$ Vm$^{9}$C$^{-5}$. The electrostrictive 
constants are given as $Q_{11}=0.11$ m$^{4}$C$^{-2}$,
$Q_{12}=-0.045$ m$^{4}$C$^{-2}$ and
$Q_{44}=0.029$ m$^{4}$C$^{-2}$.
We assume that the coefficients $g_1=g_2=g_3=g$ and use the value
$g=0.025\times{10^{-7}}$ Vm$^3$/C quoted in the literature \cite{size}. 
To calculate the rescaled quantities, we use $T_0=298K$ and $P_0=0.26$ 
 Cm$^{-2}$ and
$\delta\sim 6.7$ nm. The values chosen for the long-range parameters are
$\lambda=0.25|\alpha_1(T_0)|/{P_0}^2$ and
$\mu=20{\epsilon_0}|\alpha_1(T_0)|$.

The time-dependent Ginzburg-Landau model with the above rescaled parameters is used to simulate the domain
patterns and electromechanical properties. The equations are discretized on a
$128\times 128$ grid with the Euler scheme using periodic boundary conditions. For the length rescaling
factor $\delta \sim 6.7$ nm, this discretization corresponds to a system of size
$\sim 0.85$ $\mu$m $\times 0.85$ $\mu$m. We first simulate the properties of this $2D$ model at $T=298 K$. At this temperature,
the minima of the free energy $F_l$ define a rectangular ferroelectric phase with the four degenerate states
$(\pm 0.26, 0)$ Cm$^{-2}$ and $(0, \pm 0.26)$ Cm$^{-2}$. The dynamical equations are solved  starting from small
amplitude random
initial conditions corresponding to a quenched paraelectric phase. Domains of the four degenerate states form  and
a domain
growth process takes place. Eventually, the growth stops and a stable multi-domain state shown in
Fig. 1(a) is obtained. All four polarization variants exist in this state. Note that there are only
$90^{o}$ walls (domain walls
 across which the polarization angle changes by ninety degree) and these walls are aligned
along the [11] or [$\overline{1}$1] directions. The domain wall orientations are governed by the underlyling symmetry encoded in the
anisotropic kernel in $F_{es}$. Another interesting feature of this domain structure is that there is no
charge accumulation. This is clear from the fact that there are no head-head or tail-tail
configurations at the
domain walls.
At some of the domain junctions polarization vortices are observed and at other junctions the heads coming in
appear to balance the tails going out, thereby maintaining zero net charge.

To simulate the effect of an external electric field, the evolution
equations are solved  with a varying $E_y$ (electric field in the
$[01]$ direction) while $E_x$ is kept at zero. The domain
configuration of Fig. 1(a) is the initial condition as $E_y$ is
varied quasi-statically (the  system is allowed to relax for
$t^{*}=1000$ after each change) from $E_y=0$ to $E_y=72.07$ kV/cm
in steps of $1.85$ kV/cm. It is clear from figures 1(b)-1(d) that
this electric field causes the unfavorable domains to switch and
since the field is along one of the polarization directions,
eventually $(E_y \sim 9$ kV/cm) a single domain polarized along the
$[01]$ direction is established. On removing the field, the system
stays in this single domain state and there is an underlying
hysteresis. Figure 2(a) shows the variation of $\langle P_x
\rangle$ and $\langle P_x \rangle$ with $E_y$ for the situation
shown in Fig. 1. Also shown is the effect of electric field $E_y$
on $P_x$ and $P_y$ for an initial single domain with $P_x=0$ and
$P_y=0.26$ Cm$^{-2}$ (these single domain response curves are
obtained by minimizing the homogeneous local free energy $F_l$
only). Beyond $E_y\sim 9$ kV/cm, the single and multi-domain
responses coincide as both correspond to single domain states. 

The appearance of $\langle P_x \rangle$ during the transient indicates that the polarization reversal of the ${P_y}<0$ regions is via rotation of dipoles instead of direct $180^{o}$ flipping. Because 
the simulation size is limited, the number of clockwise and counterclockwise rotating dipoles 
is different, so that a transient $\langle P_x \rangle$ is observed. This can only be observed in 
very small 
systems in reality.
To study
the electromechanical behavior, we have also computed the behavior
of uniaxial strain with the applied electric field. The uniaxial 
strain is calculated as $\langle \eta_{yy}(E_y) \rangle -\langle
\eta_{yy}(0) \rangle$, where
$\eta_{yy}=Q_{11}{P_y}^2+Q_{12}{P_x}^2$. Figure 2(b) shows the
variation of the uniaxial strain with the electric field. The
electromechanical response of the single domain state is also
shown. It is clear that the multi-domain state generates much
higher strain compared to the single domain
case. The switching of $90^o$ domains provides the extra strain in
the multi-domain state. The piezoelectric constant $d_{33}$
(calculated as the slope of the strain vs. electric field curve)
for the large field fully polarized state is nearly the same for
both the single-domain and multi-domain responses $(\sim 82$ pC/N).
We have also calculated the dielectric constant 
$\epsilon_{yy}=1+{\epsilon_0}^{-1}(d{P_y}/d{E_y})$ for the saturated 
state from the $P_y$
vs. $E_y$ curve. The calculated dielectric constants are also
nearly the same for both single- and multi-domain states
($\epsilon_{yy}\sim 150$) . Our findings are qualitatively in agreement  
with the experimental results of Wada et al. \cite{wada} and Park et al.
\cite{epark} who have measured the electromechanical response
of $BaTiO_3$ multi-domain single crystals at room temperature. For
a quantitative comparison, a full $3D$ calculation is required.
Nevertheless, our $2D$ calculation captures the essential physics
of these experiments.

Next, we study the case with $T=273K$. At this temperature, the
free energy in $F_l$ has four minima $(\pm 0.21,\pm 0.21)$ Cm$^{-2}$
and $(\pm 0.21,\mp 0.21)$ Cm$^{-2}$, each state corresponding to a
rhombic phase. As in the earlier case, growth of domains from the
paraelectric phase results in a multi-domain state with all four
variants, as shown in Fig. 3(a). However, for this case, the
domain walls are oriented along [10] or [01] as dictated by the
symmetry of the rhombic phase encoded in the anisotropic
long-range interaction $F_{es}$. There is no net charge, even for
this case. Here also, we apply the electric field along the $[01]$
direction. This situation is interesting as the $[01]$ direction
is not one of the polarization directions at this temperature.
Because of its superior piezoelectric properties, this situation has 
been studied in numerous experimental systems 
\cite{shrout,wada, epark}.  We
vary the electric field $E_y$ at the same rate as in the $T=298K$
case. Unlike the earlier case, application of the field does not
result in transformation to a single domain state. Instead, a
stable multi-domain state that has only two of the variants $(P_y
> 0)$ is formed at $E_y=5.54$ kV/cm (Fig. 3b). This
configuration is analogous to the {\it engineered domains}
observed in recent experiments \cite{shrout, wada, epark}. This
engineered configuration persists till $E_y \sim 51 kV/cm$,
although individual polarization vectors gradually rotate towards
the [01] direction. Figure 3(c) shows the domain pattern at
$E_y=51.74$ kV/cm. Here we can clearly see that polarization
vectors have rotated and at the domain wall, the polarization is
almost aligned with the [01] direction. Thus, the domain
boundaries serve as a nucleation source for an electric field
induced structural transition from a rhombic to a rectangular
ferroelectric phase.  The transition is complete at $E_y=53.59$ 
kV/cm, as is clear from Fig. 3(d) where all the polarization
vectors are aligned along the [01] direction.  If we remove the
field, the situation shown in Fig. 3(d) remains in a single domain
rectangular state.  However, if we remove the field before the field 
induced transition, the engineered configuration remains stable.

In Fig. 4(a), we show the variation of $\langle P_x \rangle$
 and $\langle P_y \rangle$ for the situation depicted in Fig. 3. The
appropriate single domain behavior starting from a polarized state
$P_x=0.21$ Cm$^{-2}$ and $P_y=0.21$ Cm$^{-2}$ is also shown.
Polarization rotation and field induced transition at $E_y\sim
53$ kV/cm for the multi-domain state is apparent from this figure.
Interestingly, the transition occurs at a higher field  value
$E_y\sim 69$ kV/cm for the single domain state. This is due to the
fact that in the single domain state, there are no nucleation
mechanisms. Nucleation sources exist in the multi-domain state due
to the domain walls. In the single domain calculations of
Bell\cite{bell}, the field induced transitions occured at much
higher values than the experimental values. The present
calculation suggests that domain walls can help to reduce the
field level, similar to the effect of dipolar defects in reducing 
the coercive field during switching\cite{rawc}.
Figure 4(b) shows the electromechanical response for this
situation for both single and multi-domain states. These curves
are similar to the strain vs. electric field curves for experiments
where the field is applied along non-polar directions
\cite{shrout, wada, epark}. In the domain engineered regime 
(before the field induced transition), high values of $d_{33}$ due
to polarization rotations are found. For example, the multi-domain 
$d_{33}\sim 567$ pC/N and the corresonding single domain $d_{33}\sim 
367$ pC/N at $E_y=49.89$ kV/cm are achieved. The corresponding 
dielectric constants are $\epsilon_{yy}\sim 750$ (multi-domain) and 
$\epsilon_{yy}\sim 525$ (single domain). The difference between the 
single domain and multi-domain electromechanical response shows the 
property enhancement due to the presence of domain walls. 

To conclude, we have used a Ginzburg-Landau formalism to demonstrate 
the effect of the underlying domain microstructure on electromechanical 
properties of ferroelectrics. To account for nonlocal 
elastic and electrostatic effects, two additional parameters that 
measure the strength of the long-range interactions have been introduced. 
Our calculations show that these long-range parameters are essential to 
describe multi-domain states. In the present work, we have made simple 
choices for these parameters. In principle, these parameters should be 
measured for a given experimental multi-domain  state for a complete 
characterization of the material.

\newpage
\begin{figure}[h]
\includegraphics[scale=0.8]{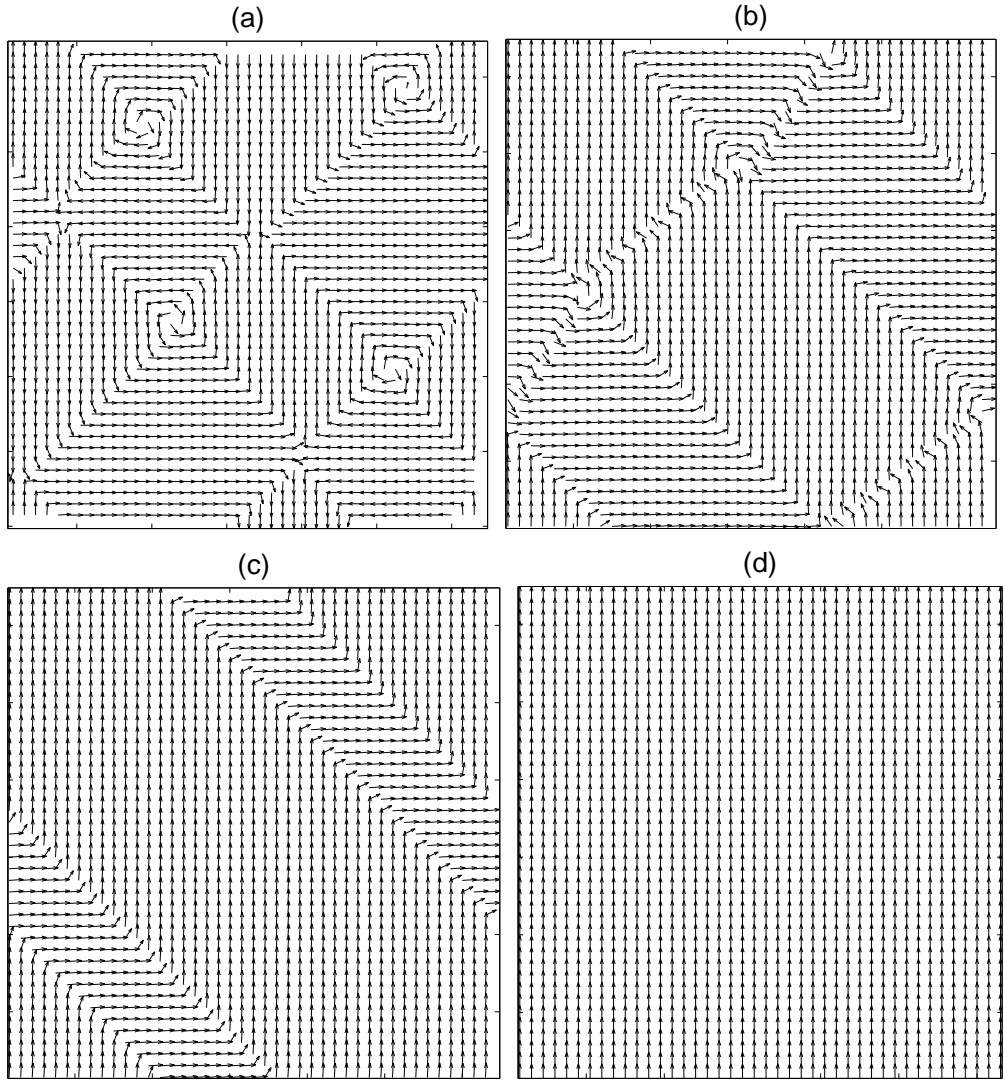}
\caption{Evolution of ferroelectric domains for a simulated ($\sim 0.85$ 
$\mu$m $\times 0.85$ $\mu$ m)
system at $T=298 K$. The  electric field values are $E_x=0, E_y=0$ (snapshot (a));
$E_x=0, E_y=5.54$ kV/cm (snapshot (b));
$E_x=0, E_y=7.39$ kV/cm (snapshot (c));
$E_x=0, E_y=9.24$ kV/cm (snapshot (d));}
\label{fig1}
\end{figure}
\newpage
\begin{figure}[h]
\includegraphics[scale=0.8]{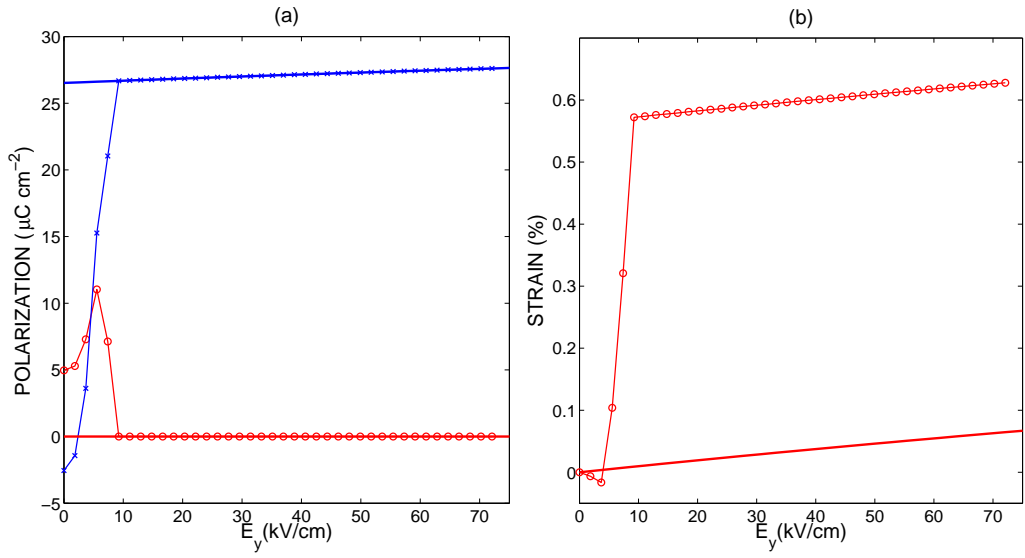}
\caption{(a) Variation of $\langle P_x \rangle$ (red line with circles)  and
$\langle P_y \rangle$ (blue line with crosses) with the applied field $E_y$ for the evolution
shown in Fig. 1. Also shown is the variation of $P_x$ (red solid line) and
$P_y$ (blue solid line)
obtained by minimizing only the free energy $F_l$ $(\lambda=\mu=0)$, corresponding to a
single domain state. (b) Variation of the uniaxial strain $\langle \eta_{yy}(E_y) \rangle
-\langle \eta_{yy}(0) \rangle$ (red line with circles) with the electric field $E_y$.
The corresponding single domain response is also shown (solid red line).}
\label{fig2}
\end{figure}
\newpage
\begin{figure}[h]
\includegraphics[scale=0.8]{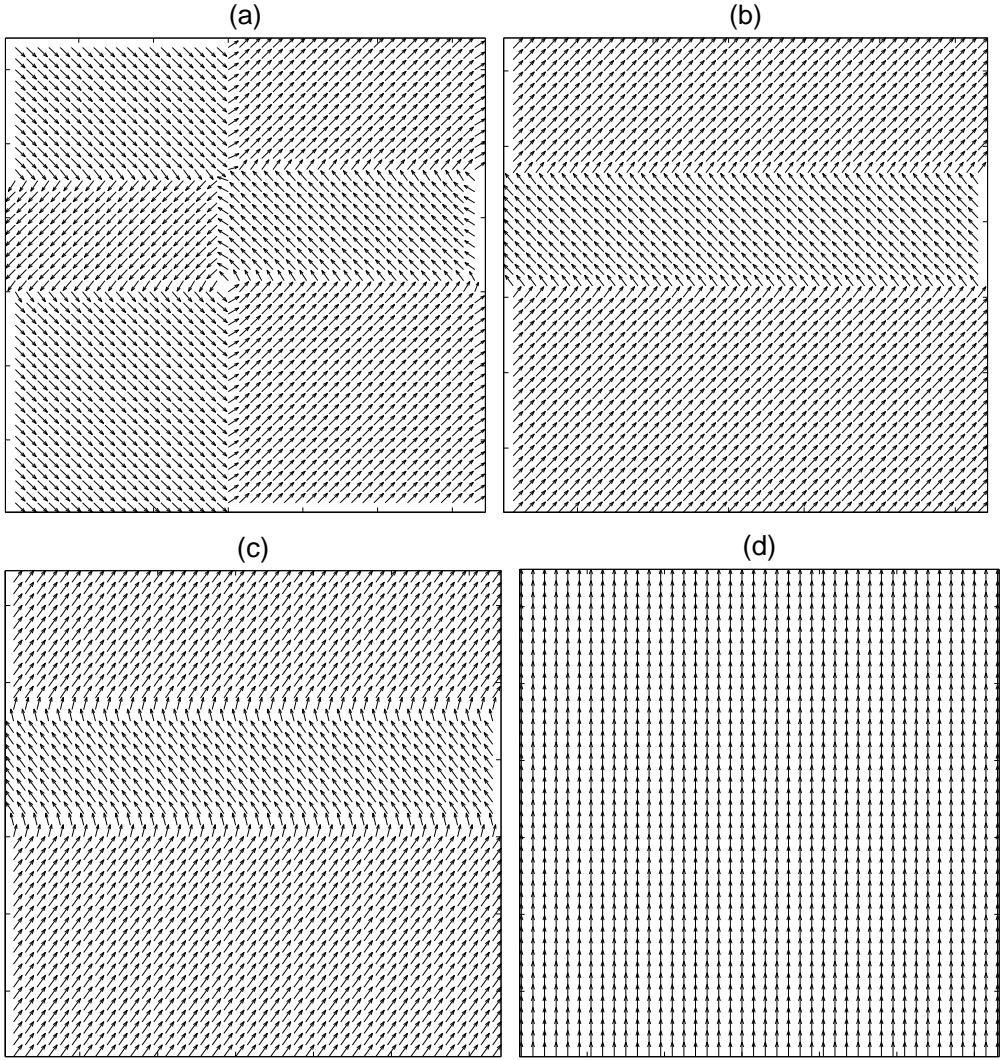}
\caption{Analogous to Fig. 1 but for $T=273 K$.
The  electric field values are $E_x=0, E_y=0$ (snapshot (a));
$E_x=0, E_y=5.54$ kV/cm (snapshot (b));
$E_x=0, E_y=51.74$ kV/cm (snapshot (c));
$E_x=0, E_y=53.59$ kV/cm (snapshot (d));}
\label{fig3}
\end{figure}
\newpage
\begin{figure}[h]
\includegraphics[scale=0.8]{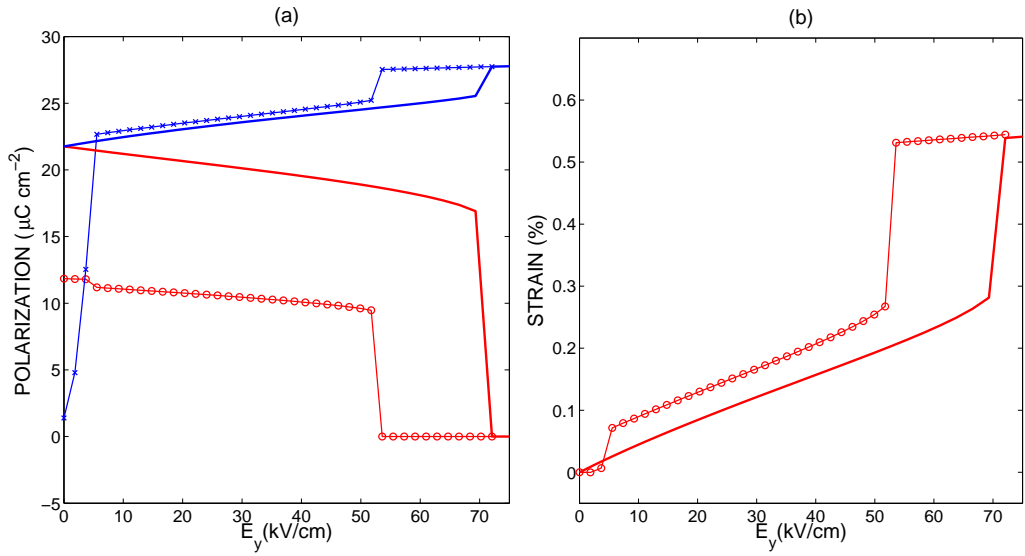}
\caption{(a) Variation of $\langle P_x \rangle$ (red line with circles)  and
$\langle P_y \rangle$ (blue line with crosses) with the applied field $E_y$ for the evolution
shown in Fig. 3. Also shown is the variation of $P_x$ (red solid line) and
$P_y$ (blue solid line) for a single domain.
(b) Variation of the uniaxial strain $\langle \eta_{yy}(E_y) \rangle
-\langle \eta_{yy}(0) \rangle$ (red line with circles) with the electric field $E_y$.
The corresponding single domain response is also shown (solid red line).}
\label{fig4}
\end{figure}
\end{document}